\documentclass[prl,showpacs,floatfix,amsmath,twocolumn,amsfonts]{revtex4}
\usepackage{amsmath,amsthm}
\usepackage{epsfig}
\usepackage{amssymb}

\begin{document}

\title{On collapse in the nonlinear Schr\"odinger equation with time dependent  nonlinearity. Application to Bose-Einstein condensates.}

\author{V. V. Konotop}
\affiliation{Centro de F\'{\i}sica Te\'orica e Computacional,
Universidade de Lisboa, Av. Prof. Gama Pinto 2, Lisboa 1649-003, Portugal}
\author{P. Pacciani}
\affiliation{Centro de F\'{\i}sica Te\'orica e Computacional, Universidade de Lisboa, Av. Prof. Gama Pinto 2, Lisboa 1649-003, Portugal}

\begin{abstract}
It is proven that periodically varying and sign definite nonlinearity in a general case does not prevent collapse in two- and three-dimensional nonlinear Schr\"{o}dinger equations: at any oscillation frequency of the nonlinearity  blowing up solutions exist. Contrary to the  results known for a sign alternating nonlinearity, increase of the frequency of oscillations accelerates collapse. The effect is discussed from the viewpoint of scaling arguments. For the three-dimensional case a sufficient condition for existence of collapse is rigorously established. The results are discussed in the context of the meanfield theory of Bose-Einstein condensates with time dependent scattering length.
\end{abstract}

 \maketitle

Stimulated by achievements in experimental realization and management of Bose-Einstein condensates (BECs), considerable attention has recently been paid to possibilities of stabilization of condensates by using time variation of the scattering length~\cite{ACKM,SaitUed,Victor,Adhikari}. By analogy with the famous Kapitza pendulum \cite{LL}, which acquires an additional dynamically stable point when the pivot is rapidly oscillating, it has been suggested~\cite{ACKM,SaitUed} that nonlinearity varying rapidly enough and changing sign during oscillations can stabilize a quasi-two-dimensional (2D) condensate. According to ~\cite{ACKM} stabilization can be achieved even when inter-atomic interactions are characterized by a negative definite scattering length.  

Previous studies were based either on qualitative arguments, like variational approach~\cite{ACKM,SaitUed,Adhikari} using the Gaussian ansatz and more accurate moment method~\cite{Victor}, or on direct numerical simulations of a multidimensional nonlinear Schr\"{o}dinger (NLS) equation. While most of the obtained results report similar conclusions about the critical collapse, approximate character of the exploited approaches results in some discrepancies in conclusions about 3D collapse. In particular, for the negative  mean (i.e. averaged over one period) scattering length the authors of Ref.~\cite{ACKM} were not able to arrest the collapse, while stable solutions were reported in Ref.~\cite{Adhikari}.
 
Thus, the present situation of the theory clearly demonstrates lack of {\em exact} results. It turns out that rigorous statements, which constitute the main goal of the present Letter, are available in a case of a sign definite scattering length~\footnote{The analogy between the problems considered in~\cite{ACKM,SaitUed} and the Kapitza pendulum is not complete: at each moment of time the pendulum has identical stability properties, while sign-alternating nonlinearity changes the stability properties of the NLS equation.}. Being subject to a number of constraints our results do not solve the problem completely, but allow one to understand the effect of the scattering length modulation on the solutions of the 2D and 3D NLS equations. In particular, we prove that variation of negative definite scattering length with any frequency generally speaking does not arrest collapse,  i.e. one can always choose an initial condition blowing up at finite time. Qualitatively, the above statement can be conjectured on the basis of earlier results. Indeed, as it has been proven in \cite{Tsutsumi}, dissipation cannot arrest the overcritical collapse, but only changes the sufficient condition of its existence. Meanwhile it is known that the varying nonlinearity in the NLS equation can be transformed into a time dependent dissipative term (see e.g.  \cite{BrazhKon}). Thus one can expect that, in a general situation, time dependent nonlinearity, even rapidly varying, will not arrest collapse, but only change conditions for this phenomenon. In this Letter, we formulate sufficient conditions for the collapse in the case of the negative mean scattering length and find that in the 3D case oscillations of the nonlinearity are favorable for the collapse, in the sense that increase of the frequency of oscillations leads to decrease of the upper bound for time of collapse.

In the case of a BEC in the mean-field approximation, the problem is described by the Gross-Pitaevskii (GP) equation~\cite{Pitaevskii}, which in the absence of the external trap potential is also known as the  NLS equation, for which the problem of collapse was intensively studied for a long time~\cite{collapse}. Taking into account that the parabolic trap potential (the typical one for the most experimental settings) does not affect the fact of the existence of collapse phenomenon in the case of a constant scattering length~\cite{Wadati}, we restrict our considerations to the NLS equation. Besides reproducing the main qualitative results for the BEC, such a statement provides generality of the results, as the NLS equation is the well known model equation for numerous physical phenomena. In particular, the results described below for the critical case are directly applicable to the problem of beam focusing in a stratified Kerr medium. 
 
{\em Statement of the problem and scaling arguments --}
Let us consider the dimensionless NLS equation
\begin{equation}
\label{NLS}
i\frac{\partial\psi_\omega}{\partial t} = -\nabla^2\psi_\omega- g(\omega t)|\psi_\omega|^{2}\psi_\omega\ ,
\end{equation}
where $\psi_\omega\equiv\psi_\omega(x,t)$ and $x\in \mathbb{R}^D$ with $D=2,3$ being the spatial dimension. The nonlinearity coefficient is considered to be varying with a period $T=2\pi/\omega$, i.e. $g(t)=g(t+2\pi)$ and to be bounded and positive definite: $g_2>g(t)>g_1>0$ for all $t$.  It will be convenient to introduce the notation $\varphi(x,t)\equiv\psi_1(x,t)$ for the solution of Eq. (\ref{NLS}) with $\omega=1$ which in this way does not contain any free parameters. Thus, $\psi_\omega(x,t)=\sqrt{\omega}\varphi\left(\omega t,\sqrt{\omega} x\right)$ is a solution of (\ref{NLS}) for a given $\omega$, whenever $\varphi(t,x)$ solves (\ref{NLS}) with $\omega=1$.
 
The energy
\begin{equation}
\label{energy_om}
E_\omega(t)=\int\left(|\nabla\psi_\omega|^2-\frac 12 g(\omega t)|\psi_\omega|^4\right)dx\ ,
\end{equation}
and the number of particles $ N_\omega=\int|\psi_\omega|^2dx$ play a special role in the analysis of the blow up phenomenon (if not specified, hereafter the integrals are taken over $\mathbb{R}^D$)~\cite{collapse}.  One easily verifies the following  relations
\begin{equation}
\label{link}
	N_\omega=\omega^{(2-D)/2}N_1\ ,\qquad E_\omega(t)=\omega^{(4-D)/2}E_1(\omega t)	\ ,
\end{equation}
The last equation, as well as the link between the solutions $\varphi$ and $\psi_\omega$, mean that existence of  a blowing up solutions of Eq. (\ref{NLS}) with $\omega=1$ and negative energy implies existence of a blowing up solution of Eq. (\ref{NLS}) at {\it any} oscillation frequency of the nonlinear term. If that happens in the critical case ($D=2$) the collapse occurs with the same number of atoms, while in the 3D case the number of particles required for the collapse decays as $1/\sqrt{\omega}$ as the frequency goes to infinity.
 
When the nonlinear term is a positive constant (i.e. when $g(t)=$const$>0$), the solution of (\ref{NLS}) blows up at a finite time, provided the energy is negative~\cite{collapse}. Below we will show that this also happens in the case of varying nonlinearity, where energy will be required to be negative at the initial moment of time. As in Eq. (\ref{link}) the link between initial energies of the  solutions $\psi_\omega$ and $\varphi$  is given by $E_\omega(0)=\omega^{(4-D)/2}E_0$ (hereafter we simplify the notation introducing $E_0=E_\omega(0)$) which means that by increasing the frequency one increases the modulus of the energy of the blowing up solution $\psi_\omega(x,t)$ proportionally to $\omega$ and $\sqrt{\omega}$ in the 2D and 3D cases respectively. 

In the case at hand, however, the energy (\ref{energy_om}) is not a constant any more but is governed by the equation
\begin{equation}
\label{en_eq}
\frac{dE_\omega}{dt}=-\frac 12 \frac{dg}{dt}\int|\psi_\omega|^4dx\ .
\end{equation}
The energy grows during half periods with $dg/dt<0$, and thus in principle may acquire positive values, even being initially negative. 
Thus the rigorous results of the NLS collapse cannot be applied straightforwardly.  
They can however be modified to provide sufficient condition for the collapse of solutions of Eq.~(\ref{NLS}), which we will discuss in the next two paragraphs (here we concentrate on physical applications of the theory; more mathematical details and generalizations will be published elsewhere).

{\em "Early-time" collapse --} Let us start with the most simple, but allowing rather general considerations, situation where $g(\omega t)$ is growing during the first half-period, and thus the energy is decaying. Then, $E_\omega(t)<0$ for the interval $t\in [0,T/2]$ (provided the solution exists) and to get a sufficient condition for the collapse it is enough to require that it happens during the first half-period (that is why we call it ``early-time'' collapse). 

This can be done by a slight modification of the standard arguments~\cite{collapse}. To this end we introduce the quantities $Y(t)=\int |x|^2|\psi_\omega|^2dx$, and $Z(t)=\mbox{Im}\int x\cdot\nabla\bar{\psi}_\omega\ \psi_\omega dx$, which solve the equations:
\begin{eqnarray}
\label{derivY}
\frac{dY(t)}{dt}&=&-4Z(t),
\\
\label{derivZ}	
\frac{dZ(t)}{dt}
&=& -DE_\omega(t)+(D-2)\int\left|\nabla\psi_\omega\right|^2dx\ .
\end{eqnarray}
From (\ref{derivY}) and (\ref{derivZ}) it follows that, if $dg/dt>0$, $E_0<0$, and $Z_0\geq 0$ (hereafter $Y(0)=Y_0$, $Z(0)=Z_0$) one can obtain the estimate $Y(t)\leq 2DE_0t^2-4Z_0t+Y_0$, from which it follows that the blow up occurs at a finite time $T_*\leq T_0<\infty$, where $\displaystyle T_0=\frac{Z_0}{DE_0}+\sqrt{\frac{Z_0^2}{D^2E_0^2}-\frac{Y_0}{2DE_0}}$. Imposing now the condition $T_0\leq T/2$ we obtain a requirement for $Y_0$: 
\begin{equation}
\label{Y0}
Y_0\leq Y_*=D|E_0|T^2/2+2Z_0T. 
\end{equation}
This condition and the requirements $E_0<0$ and $Z_0\geq 0$ constitute the sufficient conditions for the collapse to happen during the first half-period. 

The obtained result has transparent physical meaning. Indeed, compared to the standard, time independent problem, a new condition (\ref{Y0}) appeared. Since $Y(t)$ is a mean squared width of the wave packet, the new condition requires the initial wave-packet to be localized sufficiently well to decrease the blowing up time, making it less than the first half-period.

Combining the above result with the scaling arguments of the preceding paragraph, one concludes that for any oscillation frequency of the nonlinearity with initially positive derivative, one can find an initial condition for collapse at finite time, in both 2D and 3D cases. 

{\em Sufficient conditions for the collapse in the 3D case --}
Condition (\ref{Y0}) looses its practical sense in the case of rapidly varying nonlinearity, i.e. when $T\to 0$. Then, in physically relevant situations, collapse cannot occur during the first half-period and one has to consider a more general situation which will be restricted to the 3D case. Since the sign of the energy is of primary importance and assuming that initially the energy is negative, $E_0<0$, in order to establish a sufficient condition for the collapse we have to control the change of the energy $E_\omega(t)$ in time. We will do that using the ideas due to Tsutsumi~\cite{Tsutsumi}.    

Taking into account that $g(\omega t)$ is a periodic function with a period $T$, we consider an interval $t\in[T_{n-1},T_n]$, where $T_n=nT$ with $n$ being an integer, and assume that the solution exists in this interval (more precisely in the interval $t\in [0,T_n]$). As we have shown in the preceding paragraph, the way how the nonlinearity is changing during the first half-period is relevant for the early-time collapse. Now we relax this constrain, and choose the most ``unfavorable'' for collapse (because of initial grows of the energy) situation, where $dg/dt$ is time definite on each of the half-periods, with $dg/dt<0$ for $t\in (T_{n-1},T_n-T/2)$ and $dg/dt>0$ for $t\in (T_n-T/2,T_n)$. 

Next we define two functionals
\begin{eqnarray}
	\label{integrals}
	{\cal E}(t)&=&\int\left(|\nabla\psi_\omega|^2-\frac 34 g(t)|\psi_\omega|^4\right)dx\ ,
	\\
	\tilde{{\cal E}}(t)&=&\int\left(|\nabla\psi_\omega|^2-\frac 12\left(g(t)-\frac{1}{\alpha}\frac{dg}{dt}\right)|\psi_\omega|^4\right)dx.
\end{eqnarray}
Integrating by parts $e^{-\alpha t}E_\omega(t)$ with respect to time and using (\ref{en_eq}) we obtain for $t>t_1$ and for some positive constant $\alpha\geq 0$ the following relation
\begin{eqnarray}
\label{estim1}
e^{-\alpha t}E_\omega(t)=e^{-\alpha t_1}E_\omega(t_1)-\alpha\int_{t_1}^{t}e^{-\alpha s}\tilde{{\cal E}}(s) ds\ .
\end{eqnarray}

Let now $t>t_1$ and $t,\,t_1\in [T_{n-1},T_n-T/2)$. Then one has $\tilde{{\cal E}}(t_1)\leq E(t_1)$ and Eq. (\ref{estim1}) allows us to obtain
\begin{eqnarray*}
	\frac{d}{dt}\int_{t_1}^te^{\alpha (t-s)}\tilde{{\cal E}}(s)ds\leq e^{\alpha(t-t_1)}E(t_1)\ .
\end{eqnarray*}
The last formula implies 
\begin{eqnarray}
\label{estim2}
	\int_{t_1}^te^{-\alpha s}\tilde{{\cal E}}(s)ds\leq 0 \quad \mbox{if $E(t_1)<0$}. 
\end{eqnarray}
  
We have assumed that initially the energy is negative, i.e. $E_0<0$. Then, using the continuity arguments, which take into account that (\ref{estim2}) is valid for all $t$ and $t_1$ from the interval $[T_{n-1},T_n-T/2)$, we obtain that in the first half period $\tilde{{\cal E}}(t)<0$. Next we observe that $E_\omega(T/2)=\tilde{{\cal E}}(T/2)<0$ and that $E_\omega(t)$ is a decreasing function in the second half period. Hence  $E_\omega(T)<0$. Noting that $\tilde{{\cal E}}(T_{n})=E_\omega(T_{n})$ and $\tilde{{\cal E}}(T_n-T/2)=E_\omega(T_n-T/2)$ for all $n$ for which the solution exists and applying the previous arguments for the first $n$ periods, we deduce that the initial condition $E_0<0$ guarantees that $E_\omega(T_{n})<0$. In other words {\em periodically varying nonlinearity with definite sign cannot result in a change of the sign of an initially negative energy}.

For the next consideration we recall (\ref{derivY}) and (\ref{derivZ}), rewriting the  last expression for the $3D$ case as follows : $dZ/dt=-2{\cal E}(t)\geq -2\hat{\cal E}(t)$ where $\hat{\cal E}(t)$ is a continuous function defined by:
$\hat{\cal E}(t)= \tilde{{\cal E}}(t)$ when $t\in [T_{n-1},T_n-T/2]$ and  $\hat{\cal E}(t)=E_0$ when $t\in [T_n-T/2,T_n]$. 
Then the following estimate for $Y(t)$ holds
\begin{eqnarray}
\label{estim4}
	 Y(t)\leq Y_0+4\int_0^tds\left[-Z_0+2\int_0^s\hat{\cal E}(\sigma)d\sigma\right]\,.
\end{eqnarray}

Let us define $\tau=\tau(t)$ through the relation $t=nT+\tau$, where $n$ is chosen to be the largest integer assuring that $nT\leq t$ and thus $0\leq \tau<1$. Then, the first integral in (\ref{estim2}) is trivially computed, while for the second one we obtain
\begin{eqnarray*}
	 \int_0^t\int_0^s\hat{\cal E}(\sigma)d\sigma ds
	 \leq\int_0^{nT}(nT-s)\hat{\cal E}(s)ds+\left|E_0\right|T\leq\\
	 E_0T^2\left(\frac{n^2}{4}-\frac{5n}{8}\right)+\left|E_0\right|T\ .
\end{eqnarray*}

The last formula  and Eq. (\ref{estim4}) allow us to obtain the estimate as follows
\begin{eqnarray}
Y(t)\leq 2E_0T^2n^2-(5E_0T^2+4Z_0T)n+Y_0+\left|E_0\right|T
\end{eqnarray}
From this inequality  we can find the number $n_*$ determining the latest period during which blow up occurs (at that number the right hand side of the inequality becomes negative). In this way we obtain that the blow up occurs at $t<T_ {n_*}$, where
\begin{eqnarray}
\label{time}
T_*&=&\frac5 4 T+
 \frac{Z_0}{E_0}+
 \nonumber
\\
 &+& \sqrt{\frac{25}{16}T^2+\left(\frac{5Z_0}{2E_0}+\frac1 2\right)T+\frac{Z_0^2}{E_0^2}-\frac{Y_0}{2E_0}}\ .
\end{eqnarray}

Thus we have outlined the proof of the following 

THEOREM: {\em Let $\psi_\omega$ be a sufficiently smooth solution of (\ref{NLS}) in 3D case, the initial condition for which is characterized by $E_0<0$ and $Z_0\geq 0$; then blow up occurs at a finite time $t<T_*$, where $T_*$ is given by (\ref{time}).
} 
  
It is worth to emphasize that although we considered a situation where change of the scattering length is initialized with the ``negative'' half-period of $dg/dt$, the above estimates obviously applies for any initial value of  $dg/dt$.
  
{\em Estimates for real condensates --} Let us now discuss the qualitative picture emerging from the obtained results, restricting the analysis to the 3D case. We notice that the temporal characteristics of the collapse are relevant to the theory of a BEC with periodically varying scattering length due to experimental constraints on the frequency of the oscillation of the nonlinearity, emerging from the fact that change of inter-atomic interactions in practice is achieved by means of the Feshbach resonance, controlled by varying external magnetic field. The same physical phenomenon can result in creation of molecules from pairing atoms, in originating excited atomic states, etc. The respective processes are not described by the meanfield GP equation (the NLS equation), what restricts the range of meaningful frequencies. On the other hand, relevant frequencies are bounded from below by characteristic times of the condensate's life.

Although the sufficient condition gives only an upper bound for the time of the collapse, we will treat the quantities $T_0$  and $T_*$ as the collapse times (conjecturing that in a general situation decrease/increase of each of these quantities results in decrease/increase of the time of the collapse). Then the first observation is that $T_0<T_*$. Second, the upper bound for collapse $T_*$ decreases as the frequency of oscillation grows. This is in sharp contrast to what is predicted in the 2D case with the sign alternating scattering length~\cite{ACKM,SaitUed,Victor}. The third feature to be mentioned is the dependence of the collapse time not only on the number of atoms involved but on the aspect of the initial distribution. 

To connect our results with realistic experiments we provide the estimates using the data from Ref.~\cite{coll}, where observation of the collapse of a BEC controlled by Fesbach resonance (with monotonically changed magnetic field) was reported. We consider a cloud of condensed $^{85}$Rb atoms initially having a Gaussian distribution normalized to the number of particles $N$ and characterized by the radius $r$ (in dimensionless variables): $\psi(x,0)=\frac{N^{1/2}}{r^{3/2}\pi^{3/4}}\exp\left(-\frac{x^2}{2r^2}\right)$, which gives $Z_0=0$, $Y_0=3Nr^2/2$, and $E_0=\frac{3N}{2r^2}-\frac{N^2}{2^{5/2}\pi^{3/2}r^3}$. For the energy to be negative in the described situation, one must have $N>N_{cr}=2^{1/2}6\pi^{3/2}r$.

\begin{figure}[t]
\includegraphics[width=4cm,height=4cm,angle=0,clip]{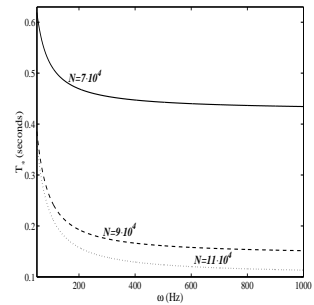} 
\caption{Upper bound for the collapse time, {\it vs} frequency for different numbers of atoms. The frequency, $\omega_0$, defined in the text for each of the curves is $4.3492\,$Hz (solid line), $13.0428\,$Hz (dashed line), and  $17.9253\,$Hz (dotted line).
} 
\label{figone}
\end{figure}

Change of the scattering length is modeled by the formula $a_s(t)=a_s^0(1-\frac{\Delta}{B(t)-B_0})$, where~\cite{feshbach} $a_s^0=-20.1\,$nm, the position of the resonance peak is $B_0=154.9\,$G and the width of the resonance is $\Delta=11\,$G. We consider the initial magnetic field $B(0)=166\,$G, which corresponds to the initial scattering length $a_s(0)=-0.18\,$nm, and the amplitude of the field oscillations $10\,$G (which for the frequency $1000\,$Hz corresponds to the speed $6.37\,$G/ms of the change of the magnetic field). Considering the initial radius of spherically symmetric cloud to be $16.5\,\mu$m (which corresponds to $r=1$ in the dimensionless units) one obtains that the link between $N$ and the real number of particles ${\cal N}$ is given by ${\cal N}\approx(N/7)\cdot 10^{4}$ (the unit of the dimensionless time corresponds to $0.116\,$s), and thus ${\cal N}$ should exceed ${\cal N}_{cr}=67498$.
In the case at hand the ``early collapse'' happens at (physical) times  bounded by $t_0\approx 13.77/\sqrt{{\cal N}-{\cal N}_{cr}}\,$s for the frequencies $\omega^2<\omega_0^2\approx 0.756\cdot 10^{-2}({\cal N}-{\cal N}_{cr})\,$ Hz. If frequency increases, or the scattering length initially decreases, collapse occurs at later times, bounded by $T_*$. Although the respective analytical expression for $T_*$ is readily obtained, it appears to be more informative to present dependence of the upper bound of the collapse time {\it vs} the frequency of the scattering length graphically (see Fig.~\ref{figone}).   

{\em Conclusion --} It has been established that nonlinearity periodic in time but sign definite does not prevent collapse in two- and three dimensional condensates with a negative mean scattering length. A sufficient condition for the collapse has been formulated which implies the possibility to create initial configurations of a condensate which will blow up in a finite time.

The sufficient condition of the 3D collapse is obviously not the optimal estimate for the time of the collapse. This is not only due to the fact that in the course of the proof some estimates were shortened (and lower precision was the price to pay for that), but mainly because the proof does not involve a specific law of the variation of the nonlinearity (but only the fact that it is periodic). The respective improvement of the estimate, as well as its generalization to the critical case (considered in the present Letter only for the case of early collapse), are left as open questions.
 
To conclude, the above results can be directly generalized to the nonlinear Shr\"{o}dinger equations with higher nonlinearity and to the NLS equation with a periodically varying dissipative term.   
 
\bigskip

We acknowledge discussions with R. Kraenkel, G. P. Menzala, J. M. Riveira, V. M. P\'erez-Garc\'{\i}a, F. Kh. Abdullaev, and S. Adhikari. 
PP was supported by the FCT fellowship SFRH/BD/16562/2004. The work was supported by the Bilateral CAPES/GRICES program.

\end{document}